\documentclass[epj-spec]{svjour}
\usepackage{graphics}
\usepackage{epsfig}
\usepackage{amssymb}
\usepackage{amsmath}
\usepackage{bm}
\begin{document}
\title{Correlated electrons in the presence of disorder}
%\subtitle{Do}
\author{K. Byczuk,\inst{1,2} W. Hofstetter,\inst{3} U. Yu,\inst{4} and D. Vollhardt\inst{2,}\thanks{\email{dieter.vollhardt@physik.uni-augsburg.de}}
%First author\inst{1}\fnmsep\thanks{\email{author@cnrs.fr}} \and Second author\inst{2} \and ... }
%
}
\institute{Institute of Theoretical Physics, University of Warsaw,
  ul. Ho\.za 69, 00-681 Warszawa, Poland
\and
Theoretical Physics III, Center for Electronic Correlations and
Magnetism, Institute of Physics, University of Augsburg, 86135
Augsburg,  Germany
\and
Institut f\"ur Theoretische Physik, Johann Wolfgang
Goethe-Universit\"at, 60438 Frankfurt/Main, Germany
\and
School of General Studies, Gwangju Institute of Science and
Technology, Gwangju 500-712, Korea
%Insert the first address here \and the second here \and ...
}
\abstract{
Several new aspects of the subtle interplay between electronic correlations and disorder are reviewed. First, the dynamical mean-field theory (DMFT)
%is employed
together with the geometrically averaged (``typical'') local density of states is employed to compute the ground state phase diagram of the Anderson-Hubbard model at half-filling. This non-perturbative approach is sensitive to Anderson localization on the one-particle level and hence can detect correlated metallic, Mott insulating and Anderson insulating phases and can also describe the competition between Anderson localization and antiferromagnetism. Second, we investigate the effect of binary alloy disorder on ferromagnetism in materials with $f$-electrons described by the periodic Anderson model. A drastic enhancement of the Curie temperature $T_c$ caused by an increase of the local $f$-moments in the presence of disordered conduction electrons is discovered and explained.
}
\maketitle
\section{Introduction}

\label{intro}

Electronic correlations are known to lead to a plethora of fascinating
phenomena \cite%
{boer37,mott37,pines62,mott90,fulde95,imada98,fazekas99,spalek00}. The same
holds true for disorder in quantum systems~\cite%
{Anderson58,LR85,VW92,KM93b,Evers08,Lagendijk09,Aspect09,Science10,Lewenstein10}%
. Both limits, electronic correlations without disorder and disorder without
electronic correlations, are notoriously difficult to treat since
interactions lead to a highly complicated many-body problem, while disorder
requires the application of statistical methods for taking averages. 
The simultaneous presence of interactions and disorder, often
encountered in real materials, therefore implies an even more complex
many-body problem \cite{Lee85,Belitz94,Kravchenko,Popovic,Kravchenko_bis,Lohneysen,ma,ma2,ma3,ma4,ma5,ma6,ma7} which is still far from
understood. Namely, repulsive interactions and disorder are both driving
forces behind metal-insulator transitions (MITs) connected with the
localization and delocalization of particles. While the Coulomb interaction
may trigger a Mott-Hubbard MIT \cite{mott90,imada98,Mott49}, the scattering
of non-interacting particles from randomly distributed impurities can lead
to Anderson localization \cite{Anderson58,Abrahams79}.

The exploration of phenomena which take place at intermediate or strong
couplings, e.g., ferromagnetism and the Mott-Hubbard MIT with and without
disorder are of particular interest. Here the recently developed dynamical
mean-field theory (DMFT) \cite{metzner89,MH,Vaclav1,Vaclav2,Georges92,Jarrell92,vollhardt93,pruschke,georges96} has
proved to provide an excellent, comprehensive mean-field approximation,
which may be employed for arbitrary values of the input parameters. For this
reason the DMFT has been successfully applied in the investigation of
electronic correlation effects in theoretical models and even real materials
\cite{georges96,Psi-k03,kotliar04,kotliar06}.

In this paper we review  our recent results on correlated electron
systems in the presence of disorder obtained within the DMFT. In particular,
we explore the low-temperature phase diagram of the Anderson-Hubbard model
at half-filling, where disordered metallic, Anderson localized, and Mott
insulating paramagnetic phases compete with antiferromagnetic long-range
order. For comparison we then investigate the phase diagram of the simpler
Falicov-Kimball model. Finally we discuss the influence of the
band-splitting by binary alloy type disorder on properties of the periodic
Anderson model. In particular, we show that alloy disorder not only leads to
a non-monotonic changes of the Curie temperature $T_{c}$ as a function of
some control parameter, and even to an enhancement of $T_{c}$ compared to
the non-disordered case, but also to the formation of Mott or Kondo
insulators at non--integer electron densities.

\section{Correlated electron systems in the presence of disorder}

The models investigated here are the Anderson-Hubbard model
\begin{equation}
H=\sum_{ij\sigma }t_{ij}c_{i\sigma }^{\dagger} c_{j\sigma}^{\phantom{+}}+
\sum_{i\sigma} \epsilon_i n_{i\sigma} + U\sum_{i}n_{i\uparrow
}n_{i\downarrow },  \label{one}
\end{equation}
where $t_{ij}$ is the hopping matrix element, $U$ is the local Coulomb
interaction, $c^{\dagger}_{i\sigma}$ ($c_{i\sigma}$) is the fermionic
creation (annihilation) operator for an
electron with spin $\sigma$ in Wannier state $i$, and $n_{i\sigma}$ is the
particle number operator; and the Anderson-Falicov-Kimball model
\begin{equation}
H=\sum_{ij}t_{ij}c^{\dagger}_i c^{}_j + \sum_i \epsilon_i c^{\dagger}_i
c^{}_i + U\sum_i f^{\dagger}_i f^{}_i c_i^{\dagger}c_i^{} ,  \label{two}
\end{equation}
where $c_i^{\dagger}$ ($f_i^{\dagger}$) and $c_i^{}$ ($f_i^{}$) are
fermionic creation and annihilation operators for \emph{mobile} (\emph{%
immobile}) particles at a lattice site $i$. Furthermore, $t_{ij}$ is the
hopping amplitude for mobile particles between sites $i$ and $j$, and $U$ is
the local interaction energy between mobile and immobile particles occupying
the same site. The ionic energy $\epsilon_i$ in both models is a random,
independent variable which describes the local, quenched disorder affecting
the motion of the mobile particles.

The disorder part is modeled by a corresponding probability distribution
function (PDF) $P(\epsilon_i)$. A model of disorder that we use 
for studying Anderson localization is one with
the continuous PDF
\begin{equation}
P(\epsilon _{i})=\frac{\Theta (\frac{\Delta}{2}-|\epsilon _{i}|)}{\Delta},
\label{four}
\end{equation}
with $\Theta $ as the step function. The parameter $\Delta $ is then a
measure of the disorder strength.

\subsection{Dynamical mean-field theory with disorder}

Dynamical mean-field theory is based on the observation that in the
high-dimensional limit $d\rightarrow \infty$ (or equivalently $Z \rightarrow
\infty$, where $Z$ is the lattice coordination number) the self-energy $%
\Sigma_{ij}(\omega)$ as defined by the Dyson equation
\begin{equation}
G_{ij\sigma}(\omega_n)^{-1}= G_{ij\sigma}^0(\omega_n)^{-1} -
\Sigma_{ij\sigma}(\omega_n),  \label{dyson}
\end{equation}
where $i,j$ denote lattice sites and $\omega_n=(2n+1)\pi/\beta$ are
fermionic Matsubara frequencies, becomes diagonal in real-space \cite%
{metzner89,MH}
\begin{equation}
\Sigma_{ij\sigma}(\omega_n)= \Sigma_{i\sigma}(\omega_n)\;\delta_{ij},
\label{local}
\end{equation}
provided the hopping amplitudes are scaled properly to ensure a balance of
kinetic and interaction energy in this limit. For nearest-neighbour hopping
on a hypercubic lattice this implies the scaling $t=t^*/\sqrt{2d}=t^*/\sqrt{Z%
}$. In a homogeneous system the self-energy is also site independent, i.e., $%
\Sigma_{ij\sigma}(i\omega_n)=\Sigma_{\sigma}(i\omega_n)\;\delta_{ij}$, and
is only a function of the energy. The DMFT approximation when applied to
finite-dimensional systems neglects off-diagonal parts of the self-energy.
In other words, the DMFT takes into account all temporal fluctuations but
neglects spatial fluctuations between different lattice sites \cite{georges96,kotliar04}.

Here we present recent developments regarding the application of DMFT to
correlated fermion systems with disorder. Within DMFT each correlated
lattice site is mapped onto a single impurity, which is coupled to a
dynamical mean-field bath describing the influence of all remaining lattice
sites. This coupling is represented by the hybridization function $%
\eta_{i\sigma}(\omega)$, which needs to be determined self-consistently. The
mapping is performed for all $N_L$ lattice sites.

In the presence of disorder as given by a particular realization of onsite
energies $\left\{ \epsilon_1, \; \epsilon_2, \; ...., \; \epsilon_{N_L}
\right\}$ the total partition function can be written as a product of the $%
N_L$ partition functions of the individual impurities
\begin{equation}
Z=\prod_{i}Z_i = \prod_{i}\exp\left( \sum_{\sigma\omega_n} \ln[%
i\omega_n+\mu-\epsilon_i - \eta_{i
\sigma}(\omega_n)-\Sigma_{i\sigma}(\omega_n) ] \right)\,.
\label{partition_bis_bis}
\end{equation}
where the hybridization function $\eta_{i \sigma}(\omega_n)$ formally
represents a site- and time-dependent one-particle potential. The unitary
time evolution due to this potential can therefore be described by a local,
time-dependent evolution operator \cite{Freericks-RMP,Freericks-book}
\begin{equation}
U[\eta_{i\sigma}]=T_{\tau} e^{-\int_0^{\beta}d\tau
\int_0^{\beta}d\tau^{\prime}c^{\dagger}_{i \sigma} (\tau)
\eta_{i\sigma}(\tau-\tau^{\prime}) c_{i\sigma}(\tau^{\prime}) }\,,
\label{evolution}
\end{equation}
where the interaction representation has been used, $T_{\tau}$ is the time
ordering operator, and $c_{i\sigma}(\tau)$ evolves according to the atomic
part $H_i^{\mathrm{loc}}$ of the Hamiltonians~(\ref{one}) or (\ref{two}) in
imaginary Matsubara time $\tau\in (0,\beta)$. We write the partition
function (\ref{partition_bis_bis}) as a trace
\begin{equation}
Z= Z [\eta_{i\sigma}]=\prod_{i=1}^{N_L}\mathrm{Tr}\left[ e^{-\beta (H_i^{%
\mathrm{loc}}-\mu N_i^{\mathrm{loc}} )} U[\eta_{i\sigma}]\right]\,,
\label{dmft}
\end{equation}
where $N_i^{\mathrm{loc}}$ is the local particle number operator.

For a given dynamical mean-field $\eta_{i \sigma}(\omega_n)$ the local
one-particle Green function $G_{ii\sigma}(\omega_n)$ is then determined from
Eq.~(\ref{dmft}) by taking a functional logarithmic derivative of the
partition function (\ref{dmft}) with respect to $\eta_{i\sigma}(\omega_n)$
\begin{equation}
G_{ii\sigma}(\omega_n)=-\frac{\partial \ln Z[\eta_{i\sigma}] }{\partial
\eta_{i\sigma}(\omega_n)}\,.  \label{local_green}
\end{equation}
We thus obtain the local Dyson equations
\begin{equation}
\Sigma_{i \sigma}(\omega_n)= i\omega_n+\mu-\epsilon_i-\eta_{i
\sigma}(\omega_n) -\frac{1}{G_{i i \sigma}(\omega_n)}\,,  \label{local_dyson}
\end{equation}
for each of the $N_L$ lattice sites. Eqs.~(\ref{dyson}, \ref{local}, \ref%
{dmft}, \ref{local_green}, and \ref{local_dyson}) constitute a closed set of
self-consistency relations. The solution of these yields an approximate
solution of the Hamiltonian~(\ref{one}) or (\ref{two}) for a given disorder
realization.

\subsection{Disorder averages}

Solving Eqs.~(\ref{dyson}, \ref{local}, \ref{dmft}, \ref{local_green}, and %
\ref{local_dyson}) rigorously is in general only possible for small $N_L$,
since it requires an exact evaluation of the time evolution operator (\ref{evolution}) at each impurity site. However, in disordered systems one is
mostly interested in the thermodynamic limit $N_L \to \infty$ where
localized and extendend states can be distinguished reliably. Here one faces
a typical trade-off situation in computational physics. A solution to this
problem is obtained by a \emph{statistical} approach in combination with
DMFT, as applied by us \cite{Byczuk05bis,Byczuk05prim,Byczuk09} and outlined
in the following.

Given a particular disorder realization $\left\{ \epsilon_1, \; \epsilon_2,
\; ...., \; \epsilon_{N_L} \right\}$ one could in principle obtain from the
solution of the DMFT equations a set of realization-dependent local
densities of states (LDOS)
\begin{equation}
A_{i\sigma}(\omega) = -\frac{1}{\pi} \mathrm{Im} G_{ii\sigma}(\omega_n%
\rightarrow \omega + i0^+)\,.  \label{spectral}
\end{equation}
Usually, however, one is interested in physical information about a system
that does not depend on a particular disorder realization. This makes a
statistical interpretation of the solutions of Eqs.~(\ref{dyson}, \ref{local}%
, \ref{dmft}, \ref{local_green}, and \ref{local_dyson}) necessary.

A common approach for systems in the thermodynamic limit $N_L\rightarrow
\infty$ is to take the arithmetic average of the LDOS $A_{i\sigma}(\omega)$
over many realizations of the disorder, i.e.,
\begin{equation}
\langle A_{i\sigma}(\omega) \rangle= \int \prod_{j=1}^{N_L}d\epsilon_j\;
P(\epsilon_i) \; A_{i\sigma}(\omega; \{\epsilon_1,...,\epsilon_{N_L}\})\,.
\label{arith}
\end{equation}
By performing the arithmetic average one restores the translational
invariance in the description of the disordered system, i.e. the average $%
A_{\sigma}(\omega)_{\mathrm{arith}} = \langle A_{i\sigma}(\omega) \rangle$
is the same for all lattice sites. However, this only leads to meaningful
results if both the system and the physical observable under study are
self-averaging.

An example of a \emph{non}-self-averaging system is a disordered system at
the Anderson localization transition
%, or a system where the localization length is larger than the diameter of the sample
\cite{Anderson58}. This implies
that during the time evolution, a particle cannot explore the full phase
space, i.e., cannot probe all possible random distributions. In such a case
the arithmetic average (\ref{arith}) is inadequate. Here one is faced with
the question concerning the proper statistical description of such a system.

As pointed out by Anderson \cite{Anderson58}, the solution to this problem
is to investigate the full PDF for a given physical observable $%
P[A_{i\sigma}(\omega)]$ and to identify its most probable value -- the
``typical'' value $A_{\sigma}(\omega)_{\mathrm{typ}}$ -- i.e. the value
where the PDF $P[A_{i\sigma}(\omega)]$ has a global maximum. This typical
value will be the same on each lattice site and represents typical properties
of the system. By considering $A_{\sigma}(\omega)_{\mathrm{typ}}$ one thus
restores translational invariance in the description of a disordered system.
Using photoemission spectroscopy one could, in principle, experimentally
probe the LDOS at different lattice sites and measure its most probable
value. We note that if sample-to-sample fluctuations are small, the typical
value $A_{\sigma}(\omega)_{\mathrm{typ}}$ coincides with the arithmetic
average $A_{\sigma}(\omega)_{\mathrm{arith}}$. On the other hand, in a
non-self-averaging system the PDF can be strongly asymmetric, with a long
tail, in which case the typical value $A_{\sigma}(\omega)_{\mathrm{typ}}$
would strongly differ from $A_{\sigma}(\omega)_{\mathrm{arith}}$. In this
case the arithmetic mean is strongly biased by rare fluctuations and hence
does not represent the typical property of such a system.

Calculating the full probability distribution of the LDOS typically requires
the inclusion of a large number of impurity sites, which for interacting
systems is hard to achieve computationally, although there have been recent
successful attempts in this direction \cite{Semmler09}. A more efficient
approach is based on identifying a generalized average which yields the best
approximation to the typical value. Among several different means the \emph{%
geometric} mean
\begin{equation}
A_{\sigma}(\omega) _{\mathrm{geom}}=\exp \left[ \langle \ln
A_{i\sigma}(\omega)\rangle \right]\,,
\end{equation}
turns out to be very convenient to describe Anderson localization, where it
represents a good approximation to the typical value:
\begin{equation}
A_{\sigma}(\omega)_{\mathrm{typ}}\approx A_{\sigma}(\omega) _{\mathrm{geom}}.
\end{equation}
Here $\langle F(\epsilon_i) \rangle =\int \prod_i d\epsilon _{i}\mathcal{P}%
(\epsilon _{i})F(\epsilon _{i})$ denotes the arithmetic mean of the function
$F(\epsilon_i)$.
It is easy to see that in the special case when  $P[A_{i\sigma}(\omega)]$ is
a log-normal distribution the relation $A_{\sigma}(\omega)_{\mathrm{typ}} =
A_{\sigma}(\omega) _{\mathrm{geom}}$ holds exactly. For the noninteracting
Anderson disorder model it was shown that $A_{\sigma}(\omega) _{\mathrm{geom}%
}$ vanishes at a critical strength of the disorder, thereby providing an
explicit criterion for Anderson localization \cite%
{Anderson58,Dobrosavljevic97,Dobrosavljevic03,Schubert03,us}. 

By using the geometrically averaged LDOS one arrives at a translationally invariant
description of the disordered system. This allows one to
solve the DMFT equations in the thermodynamic limit as will be demonstrated in the next
section. The problem of finite-size effects is then absent, and
the main limitation is due to the finite accuracy of the impurity solver.

\subsection{Self-consistency conditions of the DMFT with disorder}

Once the geometrically averaged local spectrum has been obtained, the
corresponding local Green function is given by
\begin{equation}
G_{\sigma} (\omega_n )_{\mathrm{geom}} = \int d\omega \frac{ A_{\sigma}
(\omega )_{\mathrm{geom}} }{i\omega_n-\omega }\,,
\label{local_green_average}
\end{equation}
which allows one to recast the DMFT self-consistency condition (\ref{local_dyson}%
) into the following translationally invariant form
\begin{equation}
\Sigma_{\sigma}(\omega_n)= i\omega_n+\mu-\eta_{\sigma}(\omega_n) -\frac{1}{
G_{\sigma}(\omega_n)_{\mathrm{geom}}}\,.  \label{self_invariant}
\end{equation}
where a vanishing average on-site energy $\langle \epsilon_i\rangle = 0$ has
been assumed, which holds in particular for the box-shape PDF. On the other
hand, Fourier transformation of the lattice Dyson equation (\ref{dyson})
yields
\begin{equation}
G_{\sigma} (\omega_n )_{\mathrm{geom}} = \int d z \frac{N_0(z)}{i\omega_n - z
+\mu - \Sigma_{\sigma}(\omega_n)},  \label{dyson_invariant}
\end{equation}
which closes the set of DMFT self-consistency relations. We note that the
self-consistency equation~(\ref{dyson_invariant}) gives the same geometrically
averaged Green function as that obtained from (\ref{local_green_average}).
Here $N_0(z)$ is
the density of states for a non-interacting lattice system without disorder.

The above discussion shows that by taking the geometric average of
the LDOS one i)
restores translational invariance and ii) can directly address the
thermodynamic limit. The DMFT calculations are thus no longer affected by
finite-size effects.

For the description of antiferromagnetic long-range order the
self-consistency relations need to be modified. In this case one introduces
two sublattices $s=$A or B, and calculates the two corresponding local Green
functions $G_{ii\sigma s}(\omega_n)$, which are no longer identical.
Geometric averaging of the LDOS given by Eq.~(\ref{spectral}) yields $%
A_{\sigma s}(\omega )_{\mathrm{geom}} = \exp \left[ \langle \ln A_{i\sigma
s}(\omega)\rangle \right] $. The local Green function is then again obtained
from the Hilbert transform (\ref{local_green_average}), and the self-energy $%
\Sigma _{\sigma s} (\omega )$ from the local Dyson equation Eq.~(\ref%
{self_invariant}). Finally, the self-consistent DMFT equations are closed by
the Hilbert transform of the Green function on a bipartite lattice:
\begin{equation}
G_{\sigma s}(\omega_n )_{\mathrm{geom}} =\int dz \; \frac{N_{0}(z )}{\left[%
i\omega_n -\Sigma_{\sigma s}(\omega_n )-\frac{z^2}{i\omega_n-\Sigma _{\sigma
\bar{s}}(\omega_n)}\right]}
\end{equation}
where $\bar{s}$ denotes the sublattice opposite to $s$ \cite%
{Ulmke95,georges96}.

As a concluding remark regarding the formalism we note that replacing the
geometric mean by arithmetic averaging would lead to a description of
disorder effects on the CPA level only, which would not allow to detect
Anderson localization. We also point out that in the presence of disorder
the typical LDOS $A_{\sigma}(\omega)_{\mathrm{geom}}$ is not normalized to
unity. This means that $A_{\sigma}(\omega)_{\mathrm{geom}}$ only describes
extended states, i.e. the continuum part of the spectrum. The contribution
of localized states to the LDOS is not captured by DMFT with geometric
averaging. Therefore, this approach cannot describe spectral properties of
the Anderson-insulator phase.

\section{Phase transitions in the disordered Hubbard model}

\subsection{Characterization of phases}

Our goal is to determine the ground-state phase diagram of the
Anderson-Hubbard Hamiltonian (\ref{one}). In order to characterize the
different relevant phases, we compute several physical observables:

\begin{enumerate}
\item the LDOS $A_{\sigma s}(\omega )_{\mathrm{geom}}$ for a given
sublattice $s$ and spin direction $\sigma$;

\item the total DOS for a given sublattice $s$ at the Fermi level ($\omega=0$%
) with $N_s(0)_{\mathrm{geom}}\equiv \sum_{\sigma}A_{\sigma s}(\omega=0)_{%
\mathrm{geom}}$;

\item the staggered magnetization $m_{\mathrm{AF}}^{\mathrm{geom}%
}=|n_{\uparrow A}^{\mathrm{geom}}-n_{\uparrow B}^{\mathrm{geom}}|$, where $%
n_{\sigma s}^{\mathrm{geom}}=\int_{-\infty}^0 d\omega A _{\sigma s}(\omega
)_{\mathrm{geom}}$ is the local particle density on sublattice $s$.
\end{enumerate}

The possible phases of the Anderson-Hubbard model can then be classified as
follows: The systems is a

\begin{itemize}
\item paramagnetic metal if $N_s^{\mathrm{geom}}(0)\neq 0$ and $m_{\mathrm{AF%
}}^{\mathrm{geom}}=0$;

\item AF metal if $N_s^{\mathrm{geom}}(0)\neq 0$ and $m_{\mathrm{AF}}^{%
\mathrm{geom}}\neq 0$;

\item AF insulator if $N_s^{\mathrm{geom}}(0)=0$ and $m_{\mathrm{AF}}^{%
\mathrm{geom}}\neq 0$ but $N_{ s}^{\mathrm{geom}}(\omega )\neq 0$ for some $%
\omega \neq 0$ (in fact, the last condition is already implied by $m_{%
\mathrm{AF}}^{\mathrm{geom}}\neq 0$);

\item paramagnetic Anderson-Mott insulator if $N_{s}^{\mathrm{geom}}(\omega
)=0 $ for all $\omega$.
\end{itemize}

In the following we consider the Anderson-Hubbard model at half-filling and
determine its paramagnetic as well as magnetic phases at $T=0$ using the
formalism described above \cite{Byczuk05bis,Byczuk09}. For computational
convenience, we choose a model DOS, $N_{0}(\epsilon )=2\sqrt{D^{2}-\epsilon
^{2}}/\pi D^2$, with bandwidth $W=2D$, and set $W=1$ in the following. For
this DOS and for a bipartite lattice the local Green function and the
hybridization function are connected by the simple algebraic relation $%
\eta_{\sigma s}(\omega )_{\mathrm{geom}}=D^{2}G_{\sigma \bar{s}}(\omega )_{%
\mathrm{geom}}/4$. \cite{georges96}

The resulting DMFT equations are solved at zero temperature by the numerical
renormalization group technique \cite{NRG,NRG1,NRG2}, which allows us to solve the
effective Anderson impurity problem and thus to determine the LDOS and its
averages.

\subsection{Mott-Hubbard transition \textit{vs.} Anderson localization}

\begin{figure}[tpb]
\centerline{\includegraphics[scale = .5]{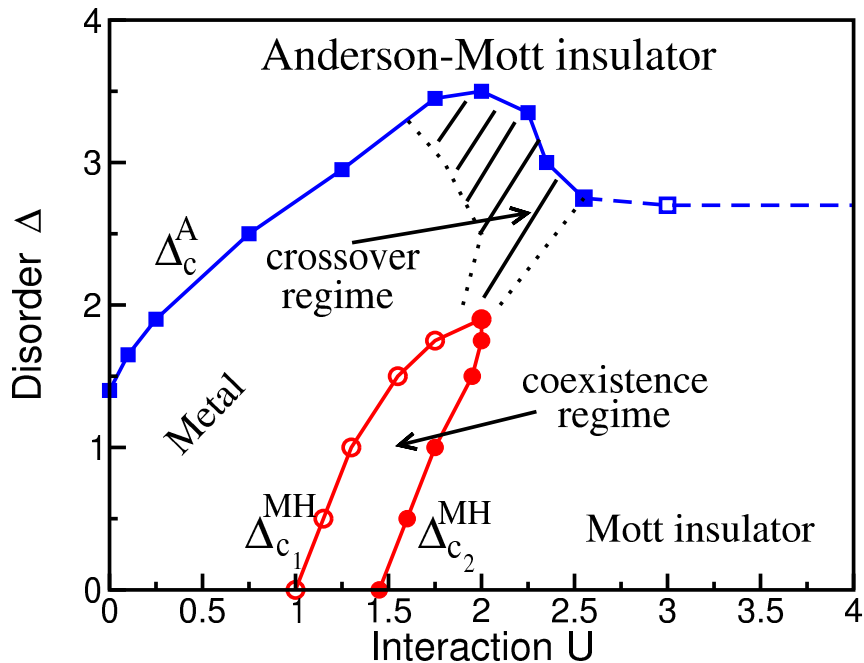}}
\caption{Paramagnetic ground-state phase diagram of the Anderson--Hubbard
model at half-filling obtained by DMFT with the typical local density of
states \protect\cite{Byczuk05bis}.}
\label{fig4.1}
\end{figure}

In Fig.~\ref{fig4.1} we show the paramagnetic ground-state phase diagram of
the Anderson-Hubbard model at half-filling obtained by the DMFT formalism
detailed above. Two different quantum phase transitions can be observed: a
Mott-Hubbard MIT for weak disorder $\Delta $, and an Anderson MIT for weak
interaction $U$. In the following we will discuss these transitions as well
as the properties of the two different insulating phases and of the
correlated, disordered metallic phase. Since these phases are all
paramagnetic the spin index will be omitted in this section.

\emph{(i) Disordered, metallic phase}: The correlated, disordered metal is
characterized by a finite value of the spectral density at the Fermi level, $%
A(\omega=0)_{\mathrm{geom}}\neq 0$.

\emph{(ii)} \emph{Mott-Hubbard MIT}: For weak to intermediate disorder
strength a sharp transition between a correlated metal and a gapped Mott
insulator occurs at a critical value of $U$. Two transition lines are found
depending on whether the MIT is approached from the metallic [$\Delta^{MH}
_{c2}(U)$, full dots in Fig. \ref{fig4.1}] or from the insulating side [$%
\Delta^{MH} _{c1}(U)$, open dots in Fig. \ref{fig4.1}]. The curves $%
\Delta^{MH}_{c1}(U)$ and $\Delta^{MH}_{c2}(U)$ in Fig. \ref{fig4.1} are seen
to have positive slope. This is due to the disorder-induced increase of
spectral weight at the Fermi level which in turn requires a stronger
interaction to open the correlation gap. As a result, close to the
hysteretic region an increase of disorder will drive the system from the
Mott insulator \emph{back} into the metallic phase or, equivalently, protect
the metal from becoming a Mott insulator.

The transition lines $\Delta^{MH}_{c1}(U)$ and $\Delta^{MH}_{c2}(U)$
terminate at a single critical point around $\Delta \approx 1.8$, cf. Fig. %
\ref{fig4.1}. For stronger disorder ($\Delta > 1.8$) we observe a smooth
crossover rather than a sharp metal-insulator transition.

\emph{(iii) Anderson MIT}: Next to the metallic phase and the crossover
regime an Anderson insulator is found where the LDOS of the extended states,
as obtained by geometric averaging, vanishes completely (see Fig. \ref%
{fig4.1}). The associated critical disorder strength $\Delta^A_{c}(U)$ at
the Anderson MIT has a non-monotonic behaviour: while it first increases as
a function of interaction in the metallic regime, it decreases again in the
crossover regime. In the former case, an increase of interaction strength
leads to re-entrant transition from the Anderson insulator into the
correlated metal, where electronic correlations prevent quasiparticles from
localization by elastic impurity scattering.

\emph{(iv) Anderson-Mott insulator}: We note that the Mott insulator (with a
finite correlation gap in the single-particle spectrum) is rigorously
defined only in the pure system ($\Delta =0$), while the gapless Anderson
insulator phase arises only for non-interacting systems ($U=0$) with
sufficiently strong disorder $\Delta >\Delta^A_{c}(0)$. In the simultaneous
presence of finite interactions and disorder it is no longer possible to
strictly distinguish these two phases. However, as long as the LDOS shows
the characteristic Hubbard subbands the system may be termed a \emph{%
disordered Mott insulator}. With increasing disorder $\Delta $ the spectral
weight of the Hubbard subbands vanishes and the system can be considered a
\emph{correlated Anderson insulator}. The (crossover) boundary between these
two types of insulators is marked by a dashed line in Fig. \ref{fig4.1}. Our
DMFT results obtained here indeed show that the paramagnetic Mott and
Anderson insulators are continuously connected. Hence, by changing $U$ and $%
\Delta$ it is possible to move from one insulating state to another one
without entering a metallic phase. This single connected, insulating phase
is therefore termed the \emph{Anderson-Mott insulator}.

\subsection{Competition between Anderson localization and antiferromagnetism}

In the previous section we have neglected antiferromagnetic ordering, which
generically occurs at low temperatures on non-frustrated lattices and in the
absence of disorder. Several fundamental questions arise: (i) How do local
interactions influence a non-interacting, Anderson localized system at
half-filling? (ii) How does an antiferromagnetic (AF) insulator at
half-filling respond to disorder which in the absence of interactions would
lead to an Anderson localized state? (iii) Can Slater and Heisenberg
antiferromagnets be distinguished by their response to disorder? Here we
provide answers to these questions by calculating the zero temperature,
magnetic phase diagram of the disordered Hubbard model at half-filling using
DMFT together with a geometric average over the disorder and allowing for a
spin-dependence of the DOS \cite{Byczuk09}.

\begin{figure}[tpb]
\centerline{\includegraphics[scale = .5]{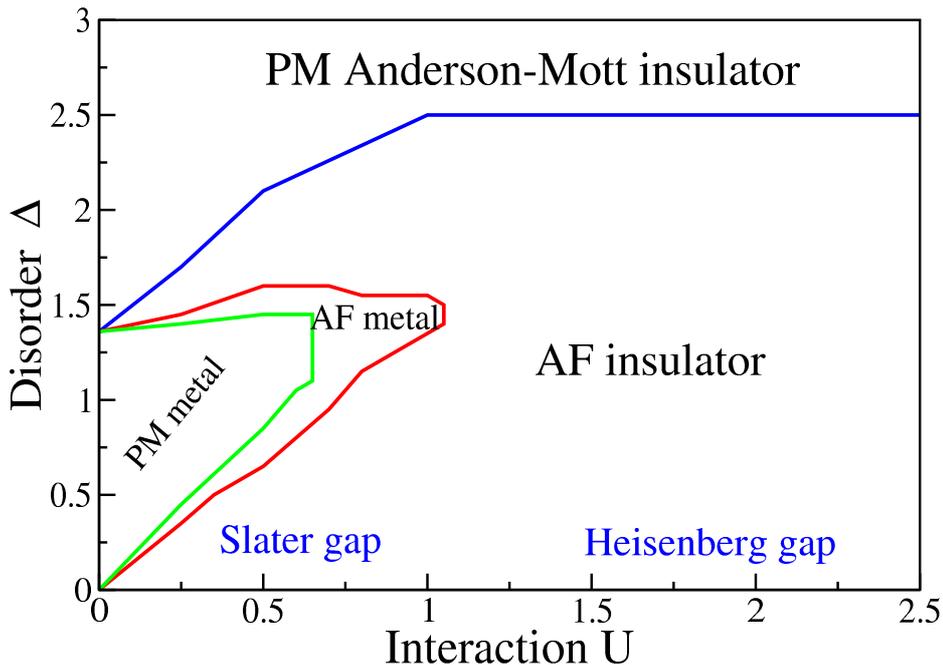}}
\caption{Magnetic ground-state phase diagram of the Anderson-Hubbard model
at half-filling as obtained by DMFT with a spin resolved, geometrically
averaged local DOS (see text) \protect\cite{Byczuk09}.}
\label{fig4.8}
\end{figure}

Our results are shown in Fig.~\ref{fig4.8}. The response of the system to
disorder is found to be qualitatively different depending on whether the
interaction $U$ is weak or strong. At strong interactions, $U/W \gtrsim 1$, only
two phases exist, an AF insulating phase at weak disorder, $\Delta /W \lesssim 2.5$%
, and a paramagnetic Anderson-Mott insulator at strong disorder, $\Delta /W
\gtrsim 2.5$. The local DOS and the staggered magnetization both decrease
gradually as the disorder $\Delta $ increases and vanish at the phase
boundary, indicating that the associated quantum phase transition is
continuous. On the other hand, a richer structure of the phase diagram is
found for weak interactions, $U/W\lesssim 1,$ (Fig.~\ref{fig4.8}). In particular,
for weak disorder a \textit{paramagnetic} metallic phase is stable. It is
separated from the AF insulator at large $U$ by a narrow region of an \textit{AF%
} \textit{metallic} phase. The AF metal is characterized by long-range order
in the absence of a gap, which is due to a redistribution of spectral weight
induced by disorder \cite{Byczuk09}.

In the pure system without disorder, the AF insulating phase can be
characterized by two limiting regimes: a Slater antiferromagnet with a
small gap at $U/W\lesssim1$ and a Heisenberg regime with large gap at
$U/W\gtrsim 1$.
These limits can be addressed by perturbation expansions in $U$ and $1/U$
around the symmetry broken state of the Hubbard and the corresponding
Heisenberg model, respectively. Our results for $m_{\mathrm{AF}}$ confirm
that even in the presence of disorder there is no sharp transition between
these limits, in agreement with earlier studies \cite{Pruschke05}. This may
be attributed to the fact that both the Slater and the Heisenberg regime are
characterized by the same order parameter. However, from our results (Fig.~%
\ref{fig4.8}) it is evident that the two limits \emph{can} be distinguished
by their overall response to disorder. Namely, the reentrance of metallic
antiferromagnetism at $\Delta/W \gtrsim  1$ occurs only within the Slater AF
insulating phase.

Let us finally remark that a disordered AF metallic phase is also obtained
by DMFT combined with arithmetic averaging \cite{Ulmke95,Singh98}. This
approach however predicts that both the paramagnetic and the AF metal remain
stable for arbitrarily strong disorder, which is clearly incorrect and
closely related to the failure of CPA to capture disorder-induced
localization. Only a computational approach which is sensitive to Anderson
localization, such as the DMFT with geometrically averaged local DOS
employed here, is able to detect already on the one-particle level the
suppression of the metallic phase for $\Delta /W\gtrsim  1.36$ and the appearance of
the paramagnetic Anderson-Mott insulator at large disorder $\Delta$.

\section{Metal-insulator transitions in the disordered Falicov-Kimball model}

We now apply the DMFT with geometric average over the disorder to the
disordered Falicov-Kimball model at half-filling \cite{Byczuk05prim}. The
non-interacting DOS for mobile particles and other physical parameters are
the same as in the previous section. However, the DMFT equations are now solved
differently: since in this case the functional integrals are Gaussian they are
calculated analytically, and only the self-consistency loops are computed
numerically.

\subsection{Ground-state phase diagram}

\begin{figure}[tb]
\centerline{\includegraphics [clip,width=12cm,angle=-0]{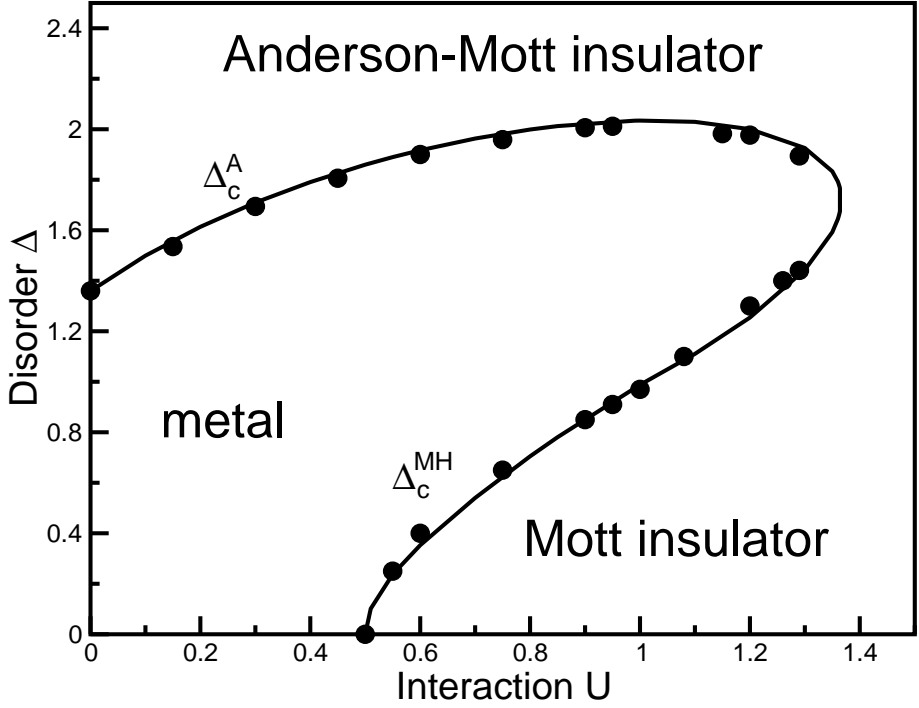}}
\nopagebreak
\caption{Ground-state phase diagram of the Falicov-Kimball model at
half-filling with disorder as obtained by the DMFT with geometric average
\protect\cite{Byczuk05prim}. Dots represent the numerical solution of the
DMFT equations. The solid line was obtained analytically by the linearized
DMFT.}
\label{Byczuk_fig3}
\end{figure}

In the half-filled case the ground-state properties of the Falicov-Kimball
model are determined by states in the center of the band, i.e., at $\omega
=0 $. The ground-state phase diagram is shown in Fig.~\ref{Byczuk_fig3}. In
the presence of disorder the metallic phase is seen to be more extended than
in the pure case. Namely, while for $\Delta =0$%
the (continuous) Mott-Hubbard MIT occurs at $U_{c}=W/2$, the presence of
disorder shifts the transition to larger values of $U$: the phase boundary
between the metal and the Mott insulator, $\Delta _{c}^{MH}(U)$ in Fig.~\ref%
{Byczuk_fig3} lies between $0\leq \Delta \lesssim 1.70W$ and $W/2\leq
U\lesssim 1.36W$. By contrast, the Anderson MIT, which is also continuous,
takes place along the phase boundary $\Delta _{c}^{A}(U)$ (see Fig.~\ref%
{Byczuk_fig3}). Again the critical disorder strength increases for
increasing $U$ and extends between $0<U\lesssim 1.36W$ and $eW/2\leq \Delta
\lesssim 2.03W$, where $e\approx 2.718$ is Euler's constant. Obviously the
interaction impedes the localization of particles due to the scattering at
the impurities. For large values of $U$ and $\Delta $ there is no metallic
phase.

The DMFT results shown here confirm that in the disordered Falicov-Kimball
model the Mott insulator and the Anderson insulator are continuously
connected. Hence, by changing $U$ and $\Delta $ it is possible to move from
one type of insulator to the other without having to cross a metallic phase.
This scenario is supported by the fact that the Anderson MIT ($U=0$) and the
Mott--Hubbard MIT ($\Delta =0$) are not associated with the breaking of a
symmetry.

\subsection{Linearized dynamical mean-field theory}

In the vicinity of the MIT the linearized DMFT \cite%
{Dobrosavljevic03,bullaldmft} may be employed to calculate the transition
line analytically. Due to the symmetry properties of $A_{\mathrm{geom}%
}(\omega )$ one finds $G(0)=-i\pi A_{\mathrm{geom}}(0)$ at the band center,
i.e., $G(0)$ is purely imaginary. The DMFT self-consistency then implies
that the hybridization function obeys $\eta (0)=-i\pi W^{2}A_{\mathrm{geom}%
}(0)/16.$%

%At this point these recursive relations turn out to be the same for arithmetic $\alpha=\rm arith $ or geometric $\alpha=\rm geom $ averaging.
Iteration of the DMFT equations leads to
\begin{equation}
A_{i}^{(n+1)}(0)=\frac{W^{2}}{16}A_{\mathrm{geom}}^{(n)}(0)\Upsilon
(\epsilon _{i}),
\end{equation}%
where
\begin{equation}
\Upsilon (\epsilon _{i})=\frac{\epsilon _{i}^{2}+\left( \frac{U}{2}\right)
^{2}}{\left[ \epsilon _{i}^{2}-\left( \frac{U}{2}\right) ^{2}\right] ^{2}}.
\end{equation}%
After the geometric average the recursive relations in the linearized DMFT
are given by
\begin{equation}
A_{\mathrm{geom}}^{(n+1)}(0)=A_{\mathrm{geom}}^{(n)}(0)\frac{W^{2}}{16}\exp %
\left[ \frac{1}{\Delta }\int_{-\Delta /2}^{\Delta /2}d\epsilon \ln \Upsilon
(\epsilon )\right] .
\end{equation}

While in the metallic phase the quantity $A_{\mathrm{geom}}^{(n)}(0)$
increases upon recursion, i.e., $A_{\mathrm{geom}}^{(n+1)}(0)>A_{\mathrm{geom%
}}^{(n)}(0)$, it decreases in the insulating phase. Therefore, at the
boundary between the metallic and the insulating solution the recursion does
not depend on $n$, i.e., $A_{\alpha }^{(n+1)}(0)=A_{\alpha }^{(n)}(0)$. This
condition directly determines the MIT transition at $\Delta =\Delta (U)$:
% shown in Fig.~\ref{Byczuk_fig3} as a solid line,
\begin{equation}
1=\frac{W^{2}}{16}\exp \left[ \frac{1}{\Delta }\int_{-\Delta /2}^{\Delta
/2}d\epsilon \ln \Upsilon (\epsilon )\right] \equiv \frac{W^{2}}{16}\exp %
\left[ I_{\mathrm{geom}}(U,\Delta )\right] .  \label{geom}
\end{equation}%
The integral is evaluated analytically with the result
\begin{eqnarray}
I_{\mathrm{geom}}(U,\Delta ) &=&2+\ln \left[ \left( \frac{U}{2}\right)
^{2}+\left( \frac{\Delta }{2}\right) ^{2}\right]  \\
&&-2\ln \left[ \left( \frac{U}{2}\right) ^{2}-\left( \frac{\Delta }{2}%
\right) ^{2}\right] +\frac{2U}{\Delta }\left[ \arctan \left( \frac{\Delta }{U%
}\right) -\ln \Big|\frac{\Delta +U}{\Delta -U}\Big|\right] .  \nonumber
\end{eqnarray}%
The solution of Eq.~(\ref{geom}), shown as a solid line in Fig.~\ref%
{Byczuk_fig3}, is found to agree very well with the numerical result. For
weak interactions $U$ the critical disorder strength obtained from (\ref%
{geom}) increases linearly in $U$, i.e., $\Delta (U)\approx We/2+\pi U/2,$
since the total bandwidth increases linearly with $U$. For weak disorder $%
\Delta $ the solution of Eq.~(\ref{geom}) is given by $\Delta (U)\approx
\sqrt{U^{2}-(W/2)^{2}}$. This agrees with the result obtained from the
arithmetic average \cite{Byczuk05prim}, which is not surprising since for
weak disorder both averages must give the same result.

\section{Ferromagnetism and Kondo insulator behavior in the disordered
periodic Anderson model}

In Secs. III and IV we discussed the results obtained for the Mott and
Anderson MITs in correlated electron systems where the disorder had a
continuous distribution, i.e., the probability distribution function of the
local atomic energies was given by a piecewise continuous function of the
energy (``box disorder''). However, to model binary-alloy systems a bimodal
probability distribution function is more appropriate. This type of disorder
also leads to a diffusive motion of the electrons, and even to Anderson
localization, as the disorder with continuously distributed energies \cite{Semmler09}. But,
for sufficiently high disorder strength the binary alloy disorder causes in
addition a splitting of the band (``alloy band-splitting'') in arbitrary
dimensions. It is therefore important to understand how the opening of two
different kinds of gaps -- the gap due to the alloy band-splitting and the
Mott gap -- affects the properties of the system.

In the following we employ the DMFT with arithmetic averaging over the
binary alloy disorder \cite{Ulmke95}. As discussed earlier, this approach
cannot detect Anderson localization on the level of one-particle quantities
such as the density of states. However, for a discussion of the effect of
band splitting on the correlated electrons such an approach is quite
appropriate. In any case, the investigation can be extended to include  the
geometric average \cite{Semmler09}.

In the case of the one-band Hubbard model with the binary-alloy disorder we
found the following interesting effects \cite{byczuk03,byczuk04}: (i) for band filling $\nu $
commensurate with the concentration of alloy atoms $x$, i.e., for $\nu =x$
or $1+x,$ the Mott-Hubbard MIT can take place for non-integer values of $\nu
$, (ii) depending on the ratio between the Mott gap and the alloy gap the
correlated alloy insulator at non-integer filling can be classified either
as a \textit{Mott-alloy insulator} (when the Mott gap is smaller than the
alloy gap) or as a \textit{charge-transfer alloy insulator} (when the Mott
gap is larger than the alloy gap), (iii) at low filling and for particular
values of the alloy concentration the Curie temperature increases in
comparison with the pure system under the same conditions. We now discuss
our results for an analogous problem, namely, the effect of binary alloy
disorder on the properties of correlated $f$ electrons which hybridize with a non-interacting band of conduction electrons, as described by the periodic
Anderson model (PAM) \cite{Yu08}.

\subsection{Periodic Anderson model with binary alloy disorder}

The Hamiltonian \ of the PAM with binary alloy disorder is given by
\begin{eqnarray}
H_{\mathrm{PAM}} &=&\sum_{ij\sigma }t_{ij}c_{i\sigma }^{\dagger }c_{j\sigma
}^{{}}+\sum_{i\sigma }\left( \varepsilon _{i}^{f}f_{i\sigma }^{\dagger
}f_{i\sigma }^{{}}+\varepsilon _{i}^{c}c_{i\sigma }^{\dagger }c_{i\sigma
}^{{}}\right)   \nonumber \\
&&+\sum_{i\sigma }\left( Vc_{i\sigma }^{\dagger }f_{i\sigma }^{{}}+V^{\ast
}f_{i\sigma }^{\dagger }c_{i\sigma }^{{}}\right) +U\sum_{i}n_{i\uparrow
}^{f}n_{i\downarrow }^{f}.  \label{4}
\end{eqnarray}%
Here $c_{i\sigma }^{\dagger }$ ($c_{i\sigma }$) and $f_{i\sigma }^{\dagger }$
($f_{i\sigma }$) are creation (annihilation) operators of conduction ($c$)
and localized ($f$) electrons with spin $\sigma $ at a lattice site $i$. The
microscopic parameters in this model are the hopping amplitude $t_{ij}$ of
the $c$-electrons, the random on--site energies $\varepsilon _{i}^{f}$ and $%
\varepsilon _{i}^{c}$, and $V$, the local hybridization between $f$- and $c$%
-electrons. The Coulomb interaction $U$ acts only between $f$-electrons on
the same site. The alloy is modeled by a bimodal probability distribution
function
\begin{equation}
P(y_{i})=x\delta (y_{i}-y_{0})+(1-x)\delta (y_{i}-y_{0}-\Delta ^{y}),
\end{equation}%
where $y_{i}=\varepsilon _{i}^{c}$, $\varepsilon _{i}^{f}$ are independent,
random variables with reference values $y_{0}=\varepsilon _{0}^{c}$, $%
\varepsilon _{0}^{f}$. The alloy concentration is characterized by the
parameter $x$ and the difference between the atomic energies of the alloy
components by $\Delta ^{y}=\Delta ^{c}$, $\Delta ^{f}$, respectively.

\subsection{Alloy-band splitting of non-interacting electrons}

The PAM [Eq.~(\ref{4})] is solved within the DMFT by mapping it onto a
corresponding single-impurity problem. The arithmetically averaged local
Green functions can be written in matrix form as%
\[
\mathbf{G}_{\sigma }^{\mathrm{loc}}(\tau ;\{y_{i}\})=-\left( \!%
%\mathcal{G}_{\sigma }^{\mathrm{loc}}(\tau ;\{y_{i}\})=-\left( \!%
\begin{array}{cc}
\langle T_{\tau }f_{\sigma }^{{}}(\tau )f_{\sigma }^{\dagger }(0)\rangle &
\langle T_{\tau }f_{\sigma }^{{}}(\tau )c_{\sigma }^{\dagger }(0)\rangle \\
\langle T_{\tau }c_{\sigma }^{{}}(\tau )f_{\sigma }^{\dagger }(0)\rangle &
\langle T_{\tau }c_{\sigma }^{{}}(\tau )c_{\sigma }^{\dagger }(0)\rangle%
\end{array}%
\!\right) .
\]%
They are expressed in terms of local self--energies which appear in the $%
\mathbf{k}$--integrated Dyson equation $\mathbf{\Sigma }_{\sigma n}=\mathbf{%
\boldsymbol{\mathcal{G}}}_{\sigma n}^{-1}-\mathbf{G}_{\sigma n}$. Here 
$\boldsymbol{\mathcal{G}}_{\sigma n}$ 
is the matrix of local Green functions of the non-interacting
bath electrons, with
\[
\boldsymbol{\mathcal{G}}_{\sigma n}^{-1}=\left(
\begin{array}{cc}
i\omega _{n}+\mu -\varepsilon _{0}^{f} & V^{\ast } \\
V & i\omega _{n}+\mu -\varepsilon _{0}^{c}-\eta _{\sigma n}%
\end{array}%
\right) .
\]

To understand the effect of the disorder on the physics described by the PAM
it is instructive to investigate the case $U=0$ first. For $U=0$ the
corresponding impurity problem is quadratic and the functional integrals can
be performed analytically. %In the
However, in the case of a two-band system like the PAM, where $f$- and $c$%
-electrons hybridize, the situation is more complicated than in the one-band
case discussed earlier, since disorder affects a hybridized two--band system
in several nontrivial ways.

\begin{figure}[t]
\epsfysize=5cm % This line scales the height of the picture
% to 3cm. You can also use \epsfxsize to set
% its width.
\centerline{\epsfbox{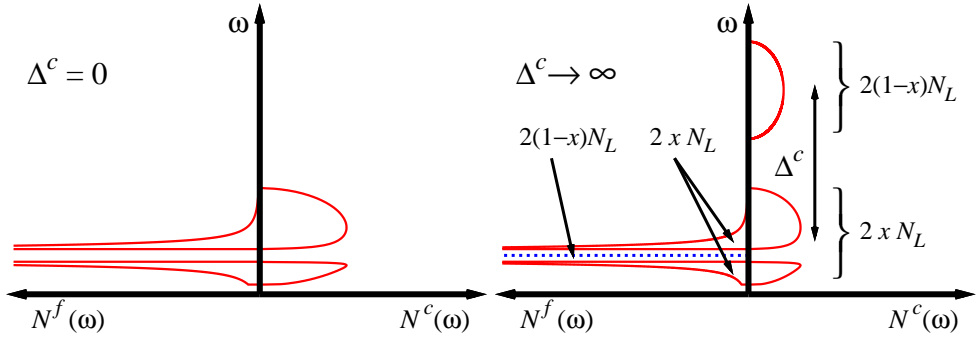}} % \centerline{} centers the picture
\caption{Binary alloy splitting of the $c$- and $f$-electron density of
states for $U=0$. a) No disorder in the $c$-electron system ($\Delta^c=0$).
b) Limit of strong disorder, $\Delta^c\rightarrow \infty$; in this case $%
2(1-x)N_L$ non-dispersive $f$-electron states (dotted line) remain at $\protect\varepsilon%
^f$ and $2(1-x)N_L$ $c$-electron states are shifted to high energies. }
\label{fig:cpa}
\end{figure}

We now consider the case where the alloy disorder acts either on the $c$%
--electrons or the $f$--electrons, %with $\varepsilon^c_i$ or
%$\varepsilon^f_i$ as random variables,
respectively. In the case of $c$-electron disorder the diagonal local Green
functions are given by
\begin{eqnarray}
G^{cc}_{\sigma n}&\!\!=\!\!& \frac{x}{(\mathcal{G}^{cc}_{\sigma n})^{-1}
-|V|^2 \mathcal{G}^{ff}_{\sigma n}} + \frac{1-x}{(\mathcal{G}^{cc}_{\sigma
n})^{-1} -|V|^2 \mathcal{G}^{ff}_{\sigma n} - \Delta^c}  \nonumber \\
G^{ff}_{\sigma n} &\!\!=\!\!& \frac{x}{(\mathcal{G}^{ff}_{\sigma n})^{-1} -
|V|^2 \mathcal{G}^{cc}_{\sigma n} } +\frac{1-x}{(\mathcal{G}^{ff}_{\sigma
n})^{-1} - \frac{|V|^2}{(\mathcal{G}^{cc}_{\sigma n})^{-1} - \Delta^c }}.
\label{diagonal}
\end{eqnarray}
The case of $f$-electron disorder is obtained by exchanging $%
f\leftrightarrow c$ in (\ref{diagonal}). Alloy disorder acting only on the $%
c $-electrons leads to a band splitting of the conduction electrons for
large enough energy splitting $\Delta^c$. As in the single--band model each
alloy subband then contains $2xN_L$ and $2(1-x)N_L$ states, respectively.
The $c$-electron alloy subbands are separated by the energy $\Delta^c$. One
might expect that, due to the hybridization of $c$- and $f$-electrons, a
similar effect would also occur in the $f$-electron subsystem. However, this
is not the case. Namely, as seen from (\ref{diagonal}) a hybridization
between the $f$-states and the $2(1-x)N_L$ states from the upper alloy $c$%
-electron subband is no longer possible for $\Delta^c\rightarrow \infty$. In
this limit a (non-dispersive) $f$-level with $2(1-x)N_L$ states appears at
the energy $\varepsilon_0^f$, in analogy with the case without hybridization
($V=0$). Consequently, for infinitely strong binary alloy disorder in the $c$%
-electron system $2(1-x)N_L$ $f$-electron states become localized for
arbitrary but finite values of $V$. So for $\Delta^c\rightarrow \infty$ only
$2(1-x)N_L$ $c$-electron states, rather than $4(1-x)N_L$ states, are split
off from the spectrum and are shifted to high energies. We note that,
although the band splitting scheme is different from the single--band model,
the alloy with hybridized $c$- and $f$-electrons can still be a band
insulator for total densities different from integer values ($2$ or $4$). A
schematic plot in Fig.~\ref{fig:cpa} shows the projected density of states, $%
N^b(\omega)=-\mathrm{Im} \sum_{\sigma} G^{bb}_{\sigma}(\omega)/\pi$, where $%
b=c$ or $f$, for a system without [panel (a)] and with [panel (b)] disorder.
An analogous analysis of $f$-electron disorder shows that in this case, at
large $\Delta^f$, the $f$-electron band is split into alloy subbands.
Hybridization between the $2(1-x)N_L$ states from the upper alloy $f$-band
and the $c$-electrons is again prevented when $\Delta^f\rightarrow \infty$.
Therefore, the corresponding fraction of the $c$-electron band is unchanged,
i.e., remains at the same energies as in the non-disordered case. We thus
see that even in the absence of interactions binary alloy disorder affects a
hybridized two--band system and a single-band system in quite different ways.

\subsection{Kondo-insulator of interacting electrons at fractional filling}

\begin{figure}[t]
\epsfxsize=9cm
\par
\centerline{ \epsfysize=7.8cm \epsfbox{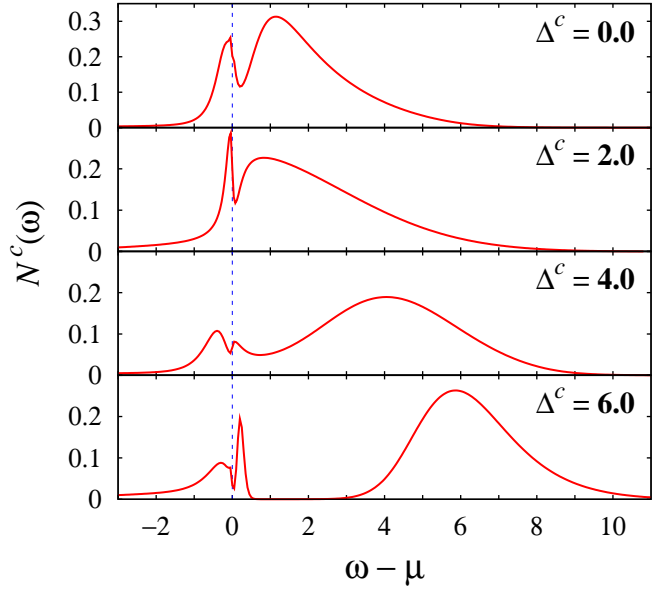}}
\caption{ Spectral function of $c$-electrons in the PAM for different $%
\Delta^c$ at $x=0.3$ (other parameters: $U=1.5$, $V=0.5$, $\protect%
\varepsilon_0^c-\protect\varepsilon^f_0=3.25 $ and $n_{\mathrm{tot}}=1.3$)
obtained within QMC and maximal entropy at $T=1/60 $. By increasing $%
\Delta^c $ a pseudogap opens, which becomes a real gap for $T \rightarrow 0$%
; after Ref. \protect\cite{Yu08}.}
\label{fig3}
\end{figure}

In analogy to our findings for the Hubbard model with binary alloy disorder
\cite{byczuk03,byczuk04} an opening of a correlation gap, obtained by
increasing the alloy band splitting $\Delta ^{c}$, is found for the PAM \cite%
{Yu08}; see Fig.~\ref{fig3}. This is caused by the splitting
of the $c$-electron band due to binary alloy disorder and the correlations
between the $f$-electrons. Namely, when the energy splitting $\Delta ^{c}$
is much larger than the width of the $c$-electron band the total number of
available low--energy states is reduced from $4N_{L}$ to $%
[4-2(1-x)]N_{L}=2(1+x)N_{L}$, whereby the filling effectively increases by a
factor of $4/[2(1+x)]$, such that $n_{\mathrm{tot}}^{\mathrm{eff}}=2n_{%
\mathrm{tot}}/(1+x)$, if $n_{\mathrm{tot}}<2(1+x)$. For the filling $n_{%
\mathrm{tot}}=1.3$ studied in Fig.~\ref{fig3}, the concentration $x=0.3$ is
a special case since then $n_{\mathrm{tot}}^{\mathrm{eff}}=2$. The system is
then effectively at half-filling and behaves as a Kondo insulator at large $%
U $, $\Delta ^{c}$, and low temperatures. The transition from a metal to a
Kondo insulator at non-integer filling predicted here for the PAM \cite{Yu08}
is a counterpart to the Mott-Hubbard metal-insulator transition at
non-integral filling in the one-band Hubbard model discussed in \cite%
{byczuk03,byczuk04}.

\subsection{Disorder induced enhancement of the Curie temperature}

It is well known that itinerant ferromagnetism occurs in the non-disordered
Hubbard model (\ref{one}) only off half-filling provided the DOS is
asymmetric and peaked at the lower edge \cite{ulmke98,wahle98}. While the
Curie temperature increases with the strength of the electron interaction
one would expect it to be lowered by disorder. However, our investigations
show that in some cases the Curie temperature can actually be increased by
binary alloy disorder \cite{byczuk03,byczuk05}.

\begin{figure}[t]
\centerline{\epsfysize=10cm \epsfbox{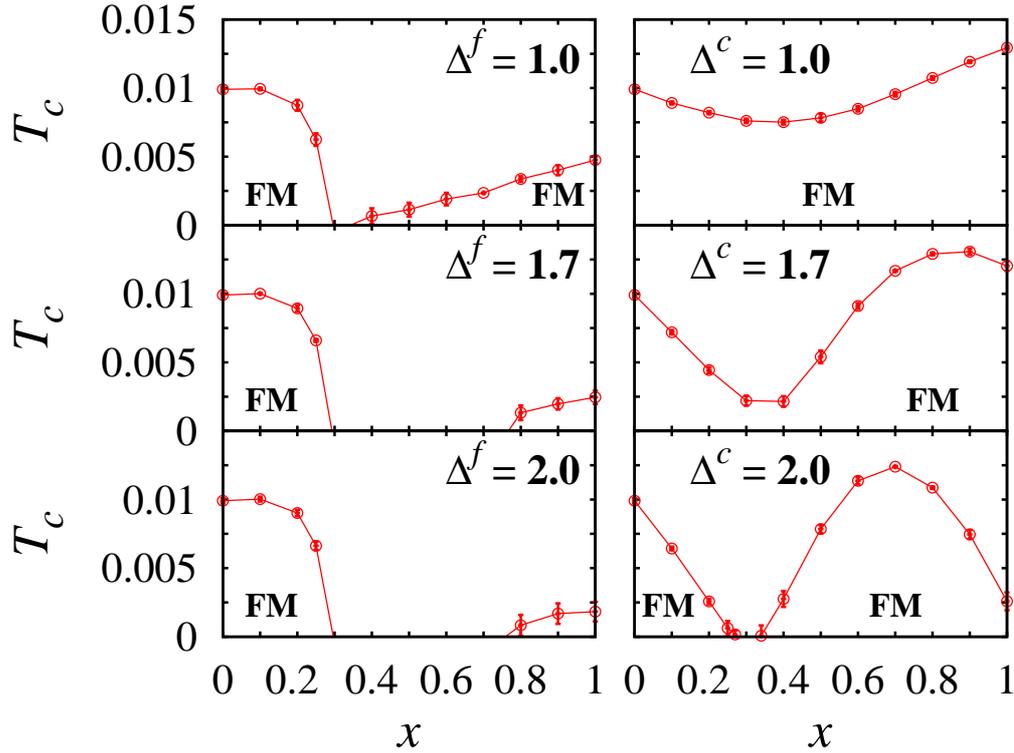}}
\caption{ Curie temperature in the PAM as a function of alloy concentration $%
x$ and energy splitting $\Delta^f$ (left column) and $\Delta^c$ (right
column) for $n_{\mathrm{tot}}=1.3$ and $\protect\varepsilon^c_0-\protect%
\varepsilon^f_0=3.25$. Strong $c$--electron disorder enhances $T_c$ compared
to its values at $x=0$ or $1$; after Ref. \protect\cite{Yu08}.}
\label{fig4}
\end{figure}

In the case of the PAM with binary alloy disorder we also found that under
certain conditions the Curie temperature can actually be enhanced \cite{Yu08}%
. Indeed, as shown in the right panel of Fig.~\ref{fig4}, the Curie
temperature for the transition to the ferromagnetic state in the PAM is a
non-monotonic function of the alloy concentration $x$. In particular, the
behavior is quite different for disorder acting on the $f$- or the $c$%
-electrons.

%\textit{$f$--electron disorder:}

\subsubsection{$f$--electron disorder}

As found by Meyer \cite{Meyer02} the presence of $f$-electron disorder
always reduces the Curie temperature relative to its non-disordered values
at $x=0$ or 1. For strong enough disorder $T_{c}$ eventually vanishes, e.g.,
at $x=0.28$ and $x=0.75$, respectively, for $\Delta ^{f}=1.7$ (right panel,
left column of Fig.~\ref{fig4}). This is due to the splitting of the $f$%
--electron band at large $\Delta ^{f}$ which increases the double occupation
of the lower alloy subband; this reduces the local moment of the $f$%
--electrons and thereby $T_{c}$.

%\textit{$c$--electron disorder:}

\subsubsection{$c$--electron disorder}

Interestingly, $c$-electron disorder leads to a much more subtle dependence
of $T_{c}$ on concentration $x$. Namely, for increasing energy splitting $%
\Delta ^{c}$ there are, in general, three different features observed, the
physical origin of which will be discussed in more detail later: \newline
(i) at $x=1$, i.e., in the non-disordered case, $T_{c}$ is reduced, \newline
(ii) a minimum develops in $T_{c}$ at $x=n_{\mathrm{tot}}-1>0$; \newline
(iii) $T_{c}$ is \emph{enhanced} over its non-disordered values at $x=0$ or $%
1$. Altogether this leads to a global maximum in $T_{c}$ vs. $x$. While the
decrease of $T_{c}$ at $x=1$ is a simple consequence of the reduction of the
energy difference between the $f$--level and the $c$-electron band, $%
\varepsilon ^{c}-\varepsilon ^{f}=\varepsilon _{0}^{c}-\varepsilon
_{0}^{f}-\Delta ^{c}$, for increasing $\Delta ^{c}$, the latter effects are
more subtle.

\subsubsection{Origin of the maximum in $T_c$}

\begin{figure}[t]
\centerline{\epsfxsize=15cm \epsfbox{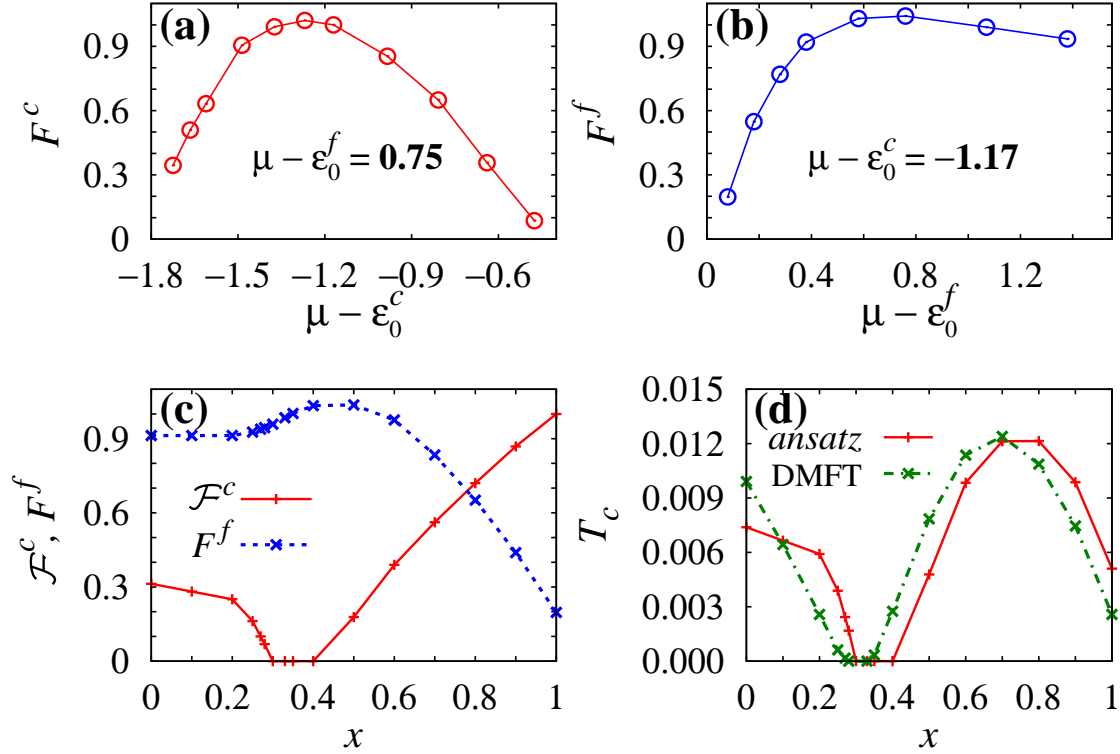} }
\caption{ (a) $F^c(\protect\mu-\protect\varepsilon^c_0)$, (b) $F^f(\protect%
\mu-\protect\varepsilon^f_0)$ appearing in the \emph{ansatz} for $T_c$ in
the disordered PAM (see text) calculated for $\Delta^c = 0$. (c) $\mathcal{F}%
^{c}(\protect\mu-\protect\varepsilon^c_0)$ and $F^f(\protect\mu-\protect%
\varepsilon^f_0)$ for $\Delta^c = 2.0$; other parameters as in Fig.~\protect
\ref{fig4}. (d) Comparison of $T_c$ obtained from the \emph{ansatz} and
within DMFT; after Ref. \protect\cite{Yu08}.}
\label{fig:scaling}
\end{figure}

We will now explain the maximum in $T_{c}$ vs. $x$. It can be understood
within the following model based on an \emph{ansatz} for the Curie
temperature, $T_{c}(U,V,\mu )=T_{c}^{0}(U,V,\mu )F^{c}(\mu -\varepsilon
_{0}^{c})F^{f}(\mu -\varepsilon _{0}^{f})$, which implies that the formation
of local $f$--electron moments ($F^{f}$) is assumed to be independent from
the $c$--electron mediated ordering of those moments ($F^{c}$). In fact, for
the RKKY model this \emph{ansatz} can be microscopically justified within a
static mean--field theory. The two functions $F^{c}$, $F^{f}$ are determined
by $T_{c}$ calculated within DMFT for the non--disorder case at fixed $\mu
-\varepsilon _{0}^{c}$ or $\mu -\varepsilon _{0}^{f}$, respectively; they
are shown in Fig.~\ref{fig:scaling}(a) and \ref{fig:scaling}(b) for one set
of parameters. The prefactor $T_{c}^{0}$ is determined by the requirement
that the dimensionless functions $F^{f}$ and $F^{c}$ be equal to one at
their maxima. We note that $F^{f}(\mu -\varepsilon _{0}^{f})$ has a maximum
when the $f$-level is half-filled ($\mu =\varepsilon _{0}^{f}+U/2$), i.e.,
when the local moment is maximal.

The Curie temperature in the presence of $c$-electron disorder can be
estimated by averaging over the $c$-electron part, $F^{c}$, giving rise to
the disorder--dependent function $\mathcal{F}^{c}(x,\mu -\varepsilon
_{0}^{c})=[xF^{c}(\mu -\varepsilon _{0}^{c}+\Delta ^{c})+(1-x)F^{c}(\mu
-\varepsilon _{0}^{c})]$. The linear dependence on the alloy concentration
can again be justified microscopically within a static mean--field theory
for the RKKY model, where $T_{c}$ depends linearly on the DOS at the
chemical potential. $T_{c}$ is now determined for each concentration $x$. We
calculate $\mu $, which is an implicit function of $x$, in the
non-hybridized limit ($V=0$) within a rigid band approximation. The
dependence of the resulting functions $\mathcal{F}^{c}(x,\mu -\varepsilon
_{0}^{c})$ and $F^{f}(\mu -\varepsilon _{0}^{f})$ on $x$ are shown in Fig.~%
\ref{fig:scaling}(c) for $\Delta ^{c}=2.0$. In general $F^{f}(\mu
-\varepsilon _{0}^{f})$ has a global maximum at those values of $x$ for
which the $f$--level is half-filled [see Fig.~\ref{fig:scaling}(c)]. By
contrast, $\mathcal{F}^{c}(x,\mu -\varepsilon _{0}^{c})$ is characterized by
a wide minimum, related to the formation of the pseudo--gap in the
interacting DOS seen in Fig.~\ref{fig3}. This minimum reaches zero, i.e., $%
\mathcal{F}^{c}(x,\mu -\varepsilon _{0}^{c})=0$, for a finite range of $x$
values as shown in Fig.~\ref{fig:scaling}(c). The resulting $T_{c}(x)$
obtained by the product of these two functions agrees remarkably well with
the numerical result obtained by DMFT as shown in Fig.~\ref{fig:scaling}(d).

\subsection{Summary}

In this paper we showed that the interplay between disorder and many-body
correlations can lead to quite unexpected, often counter-intuitive,
behavior. We first explored the zero-temperature phases of the
Anderson-Hubbard model within the dynamical mean-field theory in combination
with a geometrical average over the disorder. This allows for a unified
description of Anderson and Mott-localization on the basis of one-particle
correlation functions. The paramagnetic phase diagram shows reentrant
metal-insulator transitions caused by the interaction and disorder, and the
Anderson and Mott insulating phases were found to be continuously connected.
In the presence of antiferromagnetism a stabilizing effect of the
simultaneous presence of interaction and disorder was discovered, which
leads to a new antiferromagnetic metallic phase. An overall similar behavior
concerning Anderson and Mott insulating phases is found in the
Falicov-Kimball model. It is expected that the very interesting parameter
regime of strong interactions and strong disorder, which is not easily
accessible in correlated electron materials, can also be realized with cold
atoms in optical lattices.

In a second application we investigated ferromagnetism and Kondo insulator
behavior in the periodic Anderson model (PAM) in the presence of binary
alloy disorder. Far away from half-filling the PAM shows ferromagnetic
order. For strong enough binary alloy disorder the conduction band splits
and the correlations among the $f$-electrons lead to a non-trivial
dependence of the Curie temperature on the alloy concentration. Upon
decreasing the alloy concentration the local moments increase which raises $%
T_c$, but at the same time the opening of a gap in the alloy Kondo insulator
at non-integral filling leads to a decrease of $T_c$. In effect this causes
the Curie temperature to behave non-monotonically as a function of the alloy
concentration, with a global maximum in $T_c$ which can be drastically
larger than in the absence of disorder. The effect is predicted to occur in $%
f$-electron materials with alloy disorder in the conduction band, and also
in ultracold fermionic atoms in optical lattices trapped by harmonic
potentials in the presence of random binary disorder.

%\section*{Acknowledgements}
\medskip
{\small We thank R. Bulla and S. Kehrein for useful discussions. Financial support
by the SFB 484, TTR 80, and FOR 801 of the Deutsche Forschungsgemeinschaft is
gratefully acknowledged.}

\end{document}